\documentclass[preprint]{elsarticle}

\usepackage{graphicx}
\usepackage[caption=false,font=footnotesize]{subfig}
\usepackage{amsmath}
\usepackage{color}
\usepackage[breaklinks]{hyperref}
\usepackage{microtype}

\begin{document}

\title{Push-and-Track: Saving Infrastructure Bandwidth Through Opportunistic Forwarding}

\tnotetext[wowmom]{This article is an expanded version of work presented at the 12\textsuperscript{th} IEEE International Symposium on a World of Wireless, Mobile and Multimedia Networks (WoWMoM 2011)~\cite{Whitbeck2011}.  This paper provides greater details and presents all-new results based on different scenarios (e.g., \textit{floating data}), and different assumptions (e.g, vehicle participation ratio).}

\author[thales,lip6]{John~Whitbeck\fnref{acknowledgments}}
\ead{john.whitbeck@lip6.fr}

\author[thales]{Yoann Lopez}
\ead{yoann.lopez@fr.thalesgroup.com}

\author[thales]{J{\'e}r{\'e}mie Leguay}
\ead{jeremie.leguay@fr.thalesgroup.com}

\author[thales]{Vania~Conan}
\ead{vania.conan@fr.thalesgroup.com}

\author[lip6]{Marcelo~Dias~de~Amorim}
\ead{marcelo.amorim@lip6.fr}

\address[thales]{Thal\`es Communications\\
  160 bd de Valmy~-- 92704 Colombes Cedex~-- France}
\address[lip6]{LIP6/CNRS~-- UPMC Sorbonne Universit\'es \\
  104 avenue du Pr\'esident Kennedy~-- 75016 Paris~-- France}

\begin{abstract}
Major wireless operators are nowadays facing network capacity issues in striving to meet the growing demands of mobile users. At the same time, 3G-enabled devices increasingly benefit from ad hoc radio connectivity (e.g., WiFi). In this context of hybrid connectivity, we propose Push-and-track, a content dissemination framework that harnesses ad hoc communication opportunities to minimize the load on the wireless infrastructure while guaranteeing tight delivery delays. It achieves this through a control loop that collects user-sent acknowledgements to determine if new copies need to be re-injected into the network through the 3G interface. Push-and-Track is flexible and can be applied to a variety of scenarios, including periodic message flooding and floating data. For the former, this paper examines multiple strategies to determine \textit{how many} copies of the content should be injected, \textit{when}, and to \textit{whom}; for the latter, it examines the achievable offload ratio depending on the freshness constraints. The short delay-tolerance of common content, such as news or road traffic updates, make them suitable for such a system. Use cases with a long delay-tolerance, such as software updates, are an even better fit. Based on a realistic large-scale vehicular dataset from the city of Bologna composed of more than 10,000 vehicles, we demonstrate that Push-and-Track consistently meets its delivery objectives while reducing the use of the 3G network by about 90\%.
\end{abstract}


\maketitle

\section{Introduction}
\label{sec:introduction}

In December 2009, mobile data traffic surpassed voice on a global basis, and is expected to continue to double annually for the next five years~\cite{cisco_data,ericsson_data}. Every day, thousands of mobile devices~-- phones, tablets, cars, etc.~-- use the wireless infrastructure to retrieve content from Internet-based sources, creating immense demand on the limited spectrum of infrastructure networks, and therefore leading to deteriorating wireless quality for all subscribers as operators struggle to keep up~\cite{3g_overload}. In order to cool this surging demand, several US and European network operators have either announced or are considering the end of their unlimited 3G data plans~\cite{end_of_all_you_can_eat,3g_orange}.

There are limits however to how much can be achieved by increasing infrastructure capacity or designing better client incentives. Solving the problem of excessive load on infrastructure networks will require paradigm-altering approaches. In particular, when many users are interested in the same content, how can one leverage the multiple ad hoc networking interfaces (e.g., WiFi or Bluetooth) ubiquitous on today's mobile devices in order to assist the infrastructure in disseminating the content? Subscribers may either form a significant subset of all users, comprising for example all those interested in the digital edition of a particular newspaper, or may include all users in a given area, for example vehicles receiving periodic traffic updates in a city.

\begin{figure}
  \centering
  \includegraphics{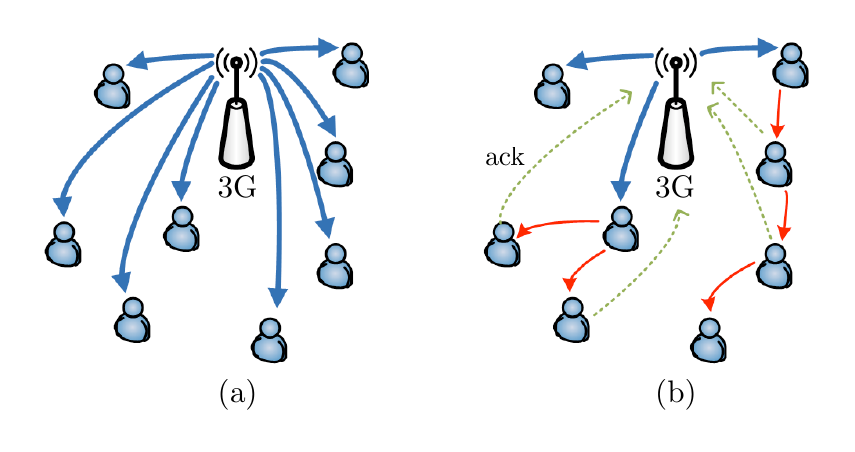}
  \caption{Combining multiple strategies for full data dissemination. Left figure (a) shows the infrastructure-only mode, where the 3G interface is used to send copies of the data to all nodes. In (b), we show the Push-and-Track approach, where opportunistic ad hoc communication is preferred whenever possible. Although acknowledgments are required to keep the loop closed, the global infrastructure load will be significantly reduced.}
  \label{fig:illu}
\end{figure}

In this paper, we address the following question: \textit{how can one relieve the wireless infrastructure using opportunistic networks while guaranteeing 100\% delivery ratio under tight delay constraints?} In particular, we seek to \textit{minimize} the infrastructure load while massively distributing content within a short time to a large number of subscribers. 

We propose \textit{Push-and-Track}, a framework that harnesses both wide-area radios (e.g., 3G or WiMax) and local-area radios (e.g., Bluetooth or WiFi) in order to achieve guaranteed delivery in an opportunistic network while relieving the infrastructure. Our approach is detailed in Fig.~\ref{fig:illu}. A subset of users will receive the content from the infrastructure and start propagating it epidemically; upon receiving the content, nodes send acknowledgments back to the source thus allowing it to keep track of the delivered content and assess the opportunity of \textit{re-injecting} copies. Since acknowledgments are assumed to be much smaller than the actual content, the load on the infrastructure would be significantly lightened. The main feature of Push-and-Track is the closed control loop that supervises the re-injection of copies of the content via the infrastructure whenever it estimates that the ad hoc mode alone will fail to achieve full dissemination within some target delay. To the best of our knowledge, our work is the first to explore this idea.

Unlike accessing an operator's wireless infrastructure, opportunistic forwarding, using short-range ad hoc radio, is essentially free and costs little more than expended battery life. This may not even be a concern in certain circumstances (e.g., vehicular). Unfortunately, it does not provide any guarantees as it depends entirely on the uncontrolled mobility of users.

In this paper, we explore strategies for two scenarios: \textit{periodic flooding}, where content must be periodically distributed to all subscribers within a maximum delay, and \textit{floating data}, where existing content must reach new nodes within a certain delay after they subscribe. Although the Push-and-Track framework may accommodate many different kinds of mobile nodes (e.g., smartphones or vehicles), this paper focuses on a vehicular scenario. All the results in this paper are based on a highly realistic large-scale vehicular simulation derived from fine-grained traffic measurements in the city of Bologna. This vehicular dataset is composed of more than 10,000 vehicles covering 20.6~km$^{2}$ and 191~km of roads. As not all nodes need necessarily participate, the above scenarios are systematically run for different percentages of participating vehicles.

In particular, in the \textit{periodic flooding} scenario, we evaluate several re-injection strategies. Push-and-Track splits the problem into \textit{how many} copies of the content should be injected into the network, \textit{when}, and to \textit{whom}. To decide the number of copies to be injected, we define different objective functions of different aggressiveness levels (slow start or fast start). If the dissemination evolution is under the objective, more copies need to be injected through the infrastructure; otherwise, the system remains in ad hoc mode only. For deciding to whom to inject copies, we consider randomized, sojourn time, location-based, and connectivity-based strategies.

We thoroughly evaluate all combinations of the proposed strategies by comparing them with both pure infrastructure and pure ad hoc approaches, as well as an oracle-based solution.

Our results reveal the following findings:

\begin{itemize}

    \item Even under tight delay constraints, Push-and-Track reduces the infrastructure load by about 90\%, when distributing periodic content to all vehicles in the city of Bologna during peak hour traffic while still achieving 100\% on time delivery ratio.

    \item Choosing random recipients for pushing content is a straightforward and efficient strategy.

    \item The feedback loop is the key feature enabling Push-and-Track to adapt to and recover from sudden network splits or varying participation rates.

\end{itemize}

In the next Section, we describe the Push-and-Track framework in more detail. In Section~\ref{sec:strategies} we detail the different scenarios and re-injection strategies tested in this paper. Our vehicular dataset is described in Section~\ref{sec:dataset}, and our simulation setup in Section~\ref{sec:setup}. The results for the periodic flooding and floating data scenarios are in Sections~\ref{sec:periodic} and~\ref{sec:floating}, respectively. Finally we discuss related work and conclude in Sections~\ref{sec:related} and~\ref{sec:conclusion}.

\section{Massive dissemination of mobile content with Push-and-Track}
\label{sec:problem}

We consider the problem of distributing dynamic content to a variable set of mobile devices, all equipped with wireless broadband connectivity (e.g., 3G) and also able to communicate in ad hoc mode (e.g., WiFi). In this paper, unless specified, we are not concerned with any specific radio technology and will simply refer to \textit{infrastructure} vs. \textit{ad hoc} radios. Mobile nodes may subscribe to various content feeds that are distributed from a point inside the infrastructure's access network and can be of any size. Whenever the subscriber base is significant enough that islands of ad hoc connectivity exist, Push-and-Track can leverage these to \emph{offload} traffic from the infrastructure to the ad hoc radio.

Push-and-Track can be thought of as ``public transportation'' for content. It shines when many mobile users are accessing the same data. The following are examples of possible use-cases:
\begin{itemize}
\item \emph{Software and system updates}. For example, tomorrow's vehicles will be equipped with many sensors and communication interfaces, all progressively integrated into Intelligent Transportation Systems (ITS)\cite{etsi}. Maintaining such a system requires each auto manufacturer to deploy a network capable of pushing software updates to entire fleets of vehicles.
\item \emph{Floating data}. Content may be of interest only in a specific geographic location. For example, a vehicle could subscribe to local traffic information upon approaching a city, and unsubscribe upon leaving it.
\item \emph{Periodic flooding}. Some news feeds or video podcasts are popular enough that updates need to be flooded to large numbers of smartphones, many of which may be within close proximity, particularly during rush hours.
\item \emph{Distributed data queries}. A centralized traffic monitoring station could, for instance, ask all vehicles in any area to issue a report on driving conditions using their sensors. Both the query and the aggregated answers could be significantly offloaded from the infrastructure to the ad hoc radio.
\end{itemize}

In this paper, all participating nodes are assumed to be interested in the content, whereas the others remain completely passive and do not participate in the epidemic dissemination. We leave the question of users forwarding content they are not interested in open for future work.

Services that are sensitive to jitter, such as VoIP, will of course remain infrastructure-only. Only content that can tolerate some delay in the delivery process (e.g., messages or file transfers) can take advantage of short range communication opportunities. Indeed, they do not have to be downloaded at the instant they are used, and can be smoothly pre-fetched into mobile devices. 

Delay-tolerance may be of two types:
\begin{enumerate}
\item \emph{Freshness-tolerance}, when the \emph{content} has a natural expiration date in terms of usefulness or validity (e.g. an hourly news update);
\item \emph{User-tolerance}, when the \emph{user} needs to receive the content within a certain time interval (e.g., receive the latest traffic update within 30 seconds of entering a city's perimeter).
\end{enumerate}
Content may have either or both kinds of delay-tolerance, which any dissemination scheme must respect.

Push-and-Track does not rely on any restricted hypothesis on contact statistics. Indeed, many opportunistic routing schemes require a learning or bootstrapping phase during which nodes aggregate statistics about meeting probabilities~\cite{lindgren03}. In particular, a lot of attention has been focused on pairwise contact and inter-contact time distributions for guiding routing decisions. In certain situations, the bootstrapping phase may be circumvented by initializing the statistics from online social network data~\cite{Bigwood2011}. Such techniques may be relevant in certain very specific circumstances, such as a conference, in which people regularly meet and separate, but are much less relevant in an urban vehicular context for example, where nodes typically meet only once. Furthermore, in a real system, users expect to be able to access the content immediately, not after some learning period. Any general realistic opportunistic content dissemination scheme which aims at guaranteeing delays cannot therefore rely only on statistical knowledge of node mobility and behavior.

In a nutshell, Push-and-Track is a mobility-agnostic framework for massively disseminating content to mobiles nodes while meeting guaranteed delays and minimizing the load on the wireless infrastructure. It consists of a control system which \textit{pushes} content to mobiles nodes and \textit{keeps track} of its opportunistic dissemination. It uses a closed-loop controller to decide when to push new copies of the content through the infrastructure and which nodes should receive them to ensure a smooth and effective dissemination using epidemic routing. Upon receiving the content, each node sends an acknowledgment back to the control system using the infrastructure network. This allows the controller to keep track of the remaining nodes to serve (see Fig.~\ref{fig:illu}). By designing the system in a way that this feedback information is much smaller than the content itself, we expect to obtain significant reduction of the traffic flowing through the infrastructure.

\section{Reducing infrastructure load: scenarios \& strategies}
\label{sec:strategies}

The content is propagating among the mobile subscribers, acknowledgments are coming in, the deadline is approaching: should copies be re-injected into the network? If so, how many and to whom? Guaranteeing 100\% delivery ratio while minimizing the load on the infrastructure is the heart of Push-and-Track. 

\subsection{Control loop operation}
\label{subsec:control-loop}

The control loop is the core of the decision system. The infrastructure must be aware of the dissemination status at all times to decide whether or not to inject new copies of the data in the network. To this end, the following control messages are mandatory. In the vehicular scenario described in the Section~\ref{sec:dataset}, each vehicle sends an \textsf{ENTER} message (i.e., subscribe) upon entering the simulation area and a \textsf{LEAVE} message (i.e., unsubscribe) upon leaving. As soon as a vehicle receives the data, it sends an \textsf{ACK} message (i.e., acknowledgment) back to the control system.

\subsection{Scenario 1: Periodic flooding}
\label{subsec:periodic_scenario}

In the periodic flooding scenario, we use Push-and-Track to guarantee the \emph{freshness} of content that has periodic updates. In this scenario, content is issued at time $t$ and must be delivered to all target nodes within a period of $T$ seconds. Nodes may enter the system in the middle of a period but they should receive the message before its expiration. Push-and-Track slots period $T$ into time steps of $\Delta_t$ seconds that correspond to the instants when the feedback loop controlling the dissemination process decides whether or not to re-inject new copies of the content. The dissemination process operates by pushing content to a subset of non infected nodes. Note that $\Delta_t$ should be greater than the transmission time of a push over the infrastructure to prevent the controller from over-reacting.

Each re-injection strategy therefore consists of two parts. At every time step, it will first determine how many, if any, copies must be re-injected, and then determine for each new copy whom to push it to.

\subsubsection{Reference strategies}
\label{subsubsec:benchmarking}

The re-injection strategies for periodic flooding will be compared to the following reference strategies:

\smallskip\noindent\textbf{Infrastructure only:} All content is pushed exclusively through the infrastructure. No ad hoc communications are allowed. This represents the baseline cost of massive content distribution using present-day deployments, and a lower limit on performance.

\smallskip\noindent\textbf{Dominating set oracle:} All content is pushed to a small number of precalculated nodes. For each message, we define a directed graph, in which each vertex is connected to all the vertices to which there exists a space-time forwarding path during the message's lifetime. Note that this directed graph does not take transmission delays into account. The infrastructure then pushes the content to a dominating set for this graph.\footnote{Here, a \textit{dominating set} is a set of nodes in the directed graph such that each node is either in the dominating set or has an inbound edge from a node in the dominating set.}  This is analogous to the well known problem of choosing multipoint relays for broadcasting in a wireless network~\cite{laouiti2001}. Finding a minimal dominating set is NP-complete but a simple greedy algorithm provides a dominating set whose cardinality is at most $\log K$ times larger than the optimal set, where $K$ is the maximum degree of a node in the aforementioned graph~\cite{laouiti2001}. As this oracle does not consider transmission times, pushing content exclusively to nodes in this dominating set would achieve near-optimal performance \textit{if ad hoc transmissions were instantaneous}. As we will see, under tight delay constraints, this oracle in fact under-estimates the number of initial copies to send.

\subsubsection{When to push}
\label{subsubsec:when}

\begin{figure}
  \centering
  \includegraphics{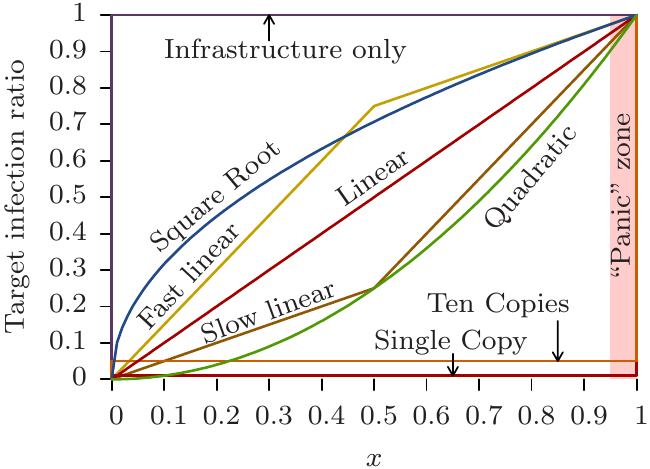}
  \caption{Infection rate objective functions. $x$ is the fraction of time elapsed between a message's creation and expiration dates. $x=1$ is the deadline for achieving 100\% infection.}
  \label{fig:objective_functions}
\end{figure}

Is it better to inject a small number of initial copies, and run the risk of having to push large numbers of copies as the deadline approaches, or jump-start the epidemic dissemination with many initial copies, despite the fact that some of those may turn out to be redundant? How about keeping a steady re-injection rate over the course of a message's lifetime? The strategies outlined in this section, hereafter called \textit{when-strategies}, cover all these questions.

Let $x$ be the fraction of time elapsed between a message's creation and expiration dates. Each strategy is defined by an \textit{objective function} (see Fig.~\ref{fig:objective_functions}), which indicates for every $0 \le x \le 1 $ what the current \textit{infection ratio} should be (i.e., the fraction of the number of subscribing nodes that have the content). Note that the infection ratio can go down if infected nodes unsubscribe or up if non-infected nodes unsubscribe. If, at any time, the measured infection ratio, obtained from the acknowledgments, is below the current target infection ratio, then the strategy returns the minimal number of additional copies that need to be re-injected in order to meet that target. Furthermore, when the time left before the deadline is equal to the time required to push the message directly through the infrastructure, the control system enters a ``panic zone'' (Fig.~\ref{fig:objective_functions}) in which the infrastructure pushes the content to all nodes that have not yet received it.

The \textit{when-strategies} may broadly be divided into three categories:

\smallskip\noindent\textbf{Slow start:} This includes two very simple ``push-and-wait'' (in opposition to Push-and-Track) strategies that push an initial number of copies and then do nothing until the panic zone: the \textit{Single Copy} and \textit{Ten Copies} strategies that respectively inject one and ten initial copies. The objective function for the \textit{Quadratic}, or ``very slow start'', strategy is $x^2$. The \textit{Slow Linear} strategy starts with a $\frac{x}{2}$ linear objective for the first half of the message's lifetime, and finishes with a $\frac{3}{2}x-\frac{1}{2}$ objective.

\smallskip\noindent\textbf{Fast start:} The objective function for the \textit{Square Root}, or ``very fast start'', strategy is $\sqrt{x}$. The \textit{Fast Linear} strategy starts with a $\frac{3}{2}x$ linear objective for the first half of the message's lifetime, and finishes with a $\frac{x}{2}+\frac{1}{2}$ objective.

\smallskip\noindent\textbf{Steady:} This is the \textit{Linear} strategy which ensures an infection ration strictly proportional to $x$.

\subsubsection{To whom}
\label{subsubsec:whom}

Once the number of copies to re-inject has been decided, the next question is whom to push it to. In this paper we test the following \textit{whom-strategies}:

\smallskip\noindent\textbf{Random:} Push to a random node chosen uniformly among those that have not yet acknowledged reception.

\smallskip\noindent\textbf{Entry time:} If content subscription is localization-based, then each node's entry time (i.e., subscription time) is correlated to its position in the area. For example, pushing to those that have the most recent (\textit{Entry-Newest}) or oldest (\textit{Entry-Oldest}) entry times should target nodes close to the edge of the area, whereas pushing to those that are closest to the average entry time (\textit{Entry-Average}) should target the middle of the area.

\smallskip\noindent\textbf{GPS-based:} On top of the existing control messages, each node may also periodically inform the control system of its current location. From this information, the space encompassing all nodes is recursively partitioned according to the Barnes-Hut method~\cite{barneshut86}. The idea is to keep on dividing each rectangular area into four sub-areas until either an area has only one node in it, or a maximum recursion level has been reached. This allows efficient computations of node density and force-based algorithms. In this paper, two GPS-based strategies were considered. In order to ensure rapid replication, \textit{GPS-Density} pushes the content to an uninfected node within the highest density area. In \textit{GPS-Potential}, each infected node $i$ applies to every other node $j$ a Coulomb potential equal to $\frac{1}{d_{ij}}$ ($d_{ij}$ is the distance between $i$ and $j$). Each side of the space also creates a potential equal to that of a single infected node. In order to spread the copies as well as possible over the entire space, \textit{GPS-Potential} pushes the content to the node with the lowest potential, i.e., the furthest away from other \emph{infected} nodes.

\smallskip\noindent\textbf{Connectivity-based:} Ad hoc routing protocols try to provide each node with a good enough picture of the global network topology to make intelligent routing decisions. On the other hand, opportunistic routing protocols only assume knowledge of the current neighbors. However, nodes can periodically communicate to the control system a list of their current neighbors. Even though each node will still perform opportunistic store-and-forwarding, the control system will have a good slightly out of sync, picture of the global connectivity graph. The \textit{CC} (Connected Components) strategy uses this information to push content to a randomly chosen node within the largest uninfected connected component. If all connected components have at least one infected node, then it pushes to a node within the one with the most uninfected nodes. The idea is to push only one copy per connected component thereby getting close to the optimal number of pushed copies.

\subsection{Scenario 2: Floating data}
\label{subsec:floating_scenario}

In the floating data scenario, we use Push-and-Track to guarantee that subscribers receive the message shortly after entering the simulation area. The floating data concept was studied by Hyytia \textit{et al.} as a way of tying content to a certain geographical area without the need for an infrastructure~\cite{Hyytia2011}. The data is replicated over ad hoc communications. Participants may enter and leave the area, but as long as enough of them are present, the content continues to ``float''. The goal is twofold: (i) ensure that the content remains present at all times in the area, and (ii) try to deliver the content to everyone passing in the area. As in most delay/disruption tolerant networking (DTN) research, success is probabilistic. Furthermore, a floating data system must cope with bringing the first initial copies or recover from content disappearance (e.g., over nighttime when an area becomes empty). 

Without a feedback loop, some nodes may never receive the content if they move through the area without coming within ad hoc transmission range of another participant; others may receive the content but only after an arbitrarily long delay. Push-and-Track may be harnessed to bring delay and reliability guarantees to floating data.

We define a \emph{user delay-tolerance} $U$. If a node subscribes to the content at time $t_i$, then it must have received the content before time $t_i + U$. Unlike the periodic flooding scenario, we are not interested in the dynamics of content dissemination and will start the scenario in a state where the content is already widely disseminated. The focus of this scenario is on the time it takes for newly arrived nodes to receive the content epidemically. The aim is to determine how much data may be offloaded from the 3G infrastructure while fulfilling user delay-tolerance.

As in the periodic flooding scenario, the reference strategy is the one that exclusively uses the infrastructure and does not allow any ad hoc communications. This represents the baseline cost of pushing content to nodes entering an area using current deployments. The Push-and-Track strategy in this case is very simple. As previously, thanks to the \textsf{ENTER}, \textsf{LEAVE}, and \textsf{ACK} control messages, the controller can track who has or has not received the content. If a node has subscribed to the content, but has not received it from ad hoc dissemination within time $U$, the controller pushes the content through the infrastructure.

\section{Bologna vehicular dataset}
\label{sec:dataset}

Many existing datasets were considered for evaluating Push-and-Track, in particular the Bluetooth contact traces obtained in a conference~\cite{chaintreau}, on a campus~\cite{mit}, or during a rollerblading tour~\cite{tournoux08}. Unfortunately, these all have a small fixed set of participants (roughly 100) and the underlying social affinities and dynamics translate into specific inter-contact patterns that have a crucial impact on data dissemination. For our purposes, we wanted a realistic dataset with a large variable number of users and a high turnover rate among the users to simulate subscription and unsubscription. Furthermore, as in real-life, we expect those users to be mostly strangers to each other, and therefore wished to keep social dynamics to a minimum. The Bologna vehicular dataset described in this section has all these features.

\subsection{Dataset construction}
\label{subsec:dataset_construction}

\begin{figure}
    \centering
    \includegraphics[scale=0.5]{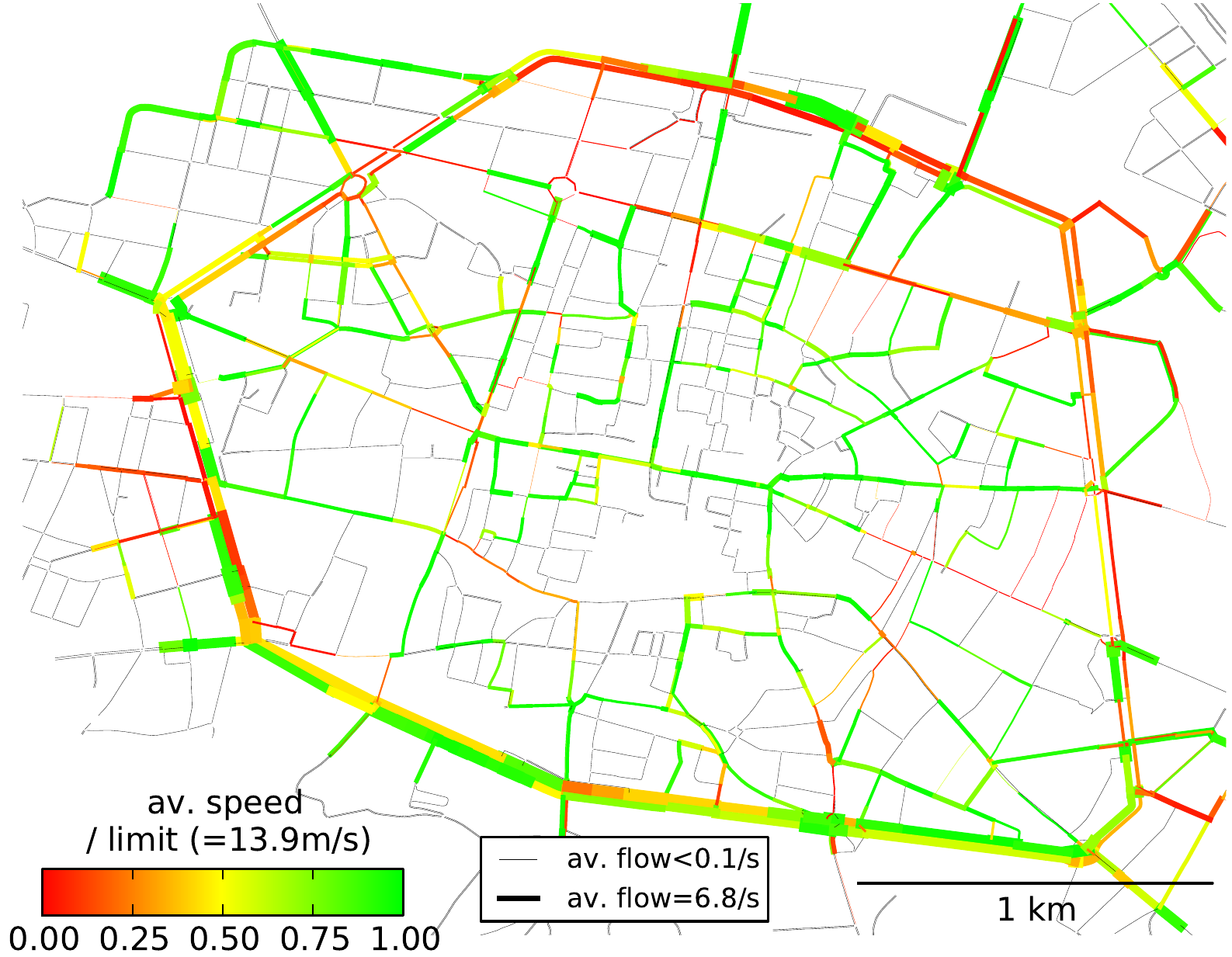}
    \caption{Simulated Bologna road network. The incoming vehicle flow on a given lane is anamorphically represented by its width on a logarithmic scale. The vehicle speed on a lane is represented by a color gradient. Both values are averaged over the duration of the dataset.}
    \label{fig_map}
\end{figure}

We evaluate Push-and-Track on a large-scale vehicular mobility excerpt of a city-wide dataset of the municipality of Bologna (Italy). This dataset's initial purpose was to evaluate future cooperative road traffic management strategies within the iTetris European project and is available for download from the project's website~\cite{itetris}. In this paper, we focused on the area surrounding Bologna's city center, displayed in Fig.~\ref{fig_map}, covering $20.6$ km$^2$ and including $191$ km of roads. It exhibits diversity in terms of road types: a ring-shaped main road yields to various entry points to the historical city center.

The dataset is derived from measurements of traffic conditions realized by the municipality of Bologna on their road network. The supplied raw data includes measurements of circulating vehicles acquired by 636 induction loops spread over the city and a synthesis of user surveys on usual commuting trips. Exploiting this raw data, the OD (Origin--Destination) vehicle flow matrices yield macroscopic traffic demands on the city road network during common weekday peak hours (from 8~a.m. to 9~a.m.). Monday and Friday mornings were discarded to avoid specific traffic patterns due to week-end proximity.

Using common traffic engineering tools~\cite{vissim}, the macroscopic traffic demands and route assignments are then used to infer individual vehicle micro-mobility on a highly-accurate representation of the Bologna road network. We ran the simulation with SUMO, an open-source microscopic vehicular movement simulator generally used by the vehicular research community for testing and comparing models of vehicle behavior, traffic light optimization, and vehicle routing~\cite{krajzewicz2002sumo}. To model individual vehicle behavior, SUMO uses a space-continuous and time-discrete car-following model on a multiple-lane road network representation~\cite{zpr98-319}. The latter is supplied in the Bologna dataset and includes traffic lights' positions and patterns, lane-changing, and junction-based right-of-way rules.

\subsection{Dataset analysis}
\label{subsec:dataset_analysis}

We now analyze the vehicular traffic and network connectivity statistics of the simulation. After a warm-up period, the traffic is simulated during 3,600~s, which leads to a total number of 10,333 simulated vehicles. During this hour, a maximum of 4,494 and an average of 3,540 vehicles are simultaneously present on the road network. As in real-life, traffic conditions vary from fluid to congested in different parts of the city. This is reflected in the vehicles' transit times. Indeed, vehicles remain an average of 13.2 minutes in the city area. While most of these are short trips (50\% are below 10 minutes), some last for over 50 minutes long. Fig.~\ref{fig_map} shows the number of vehicles and average speeds on each road in Bologna. It highlights the relatively larger amount of traffic on the surrounding ring-shaped multiple-lane road than on the capillary network, which is mainly single-lane. Due to dense morning traffic, right-of-way rules, and traffic lights, traffic jams occur on the outer ring and at crossroads.

\begin{table}
  \centering
  \caption{Characteristics of the dataset for different percentages of participating vehicles. Number of vehicles, connected components, singletons, as well as connected component size and node degree are weighted averages over each dataset's duration}.
  \includegraphics{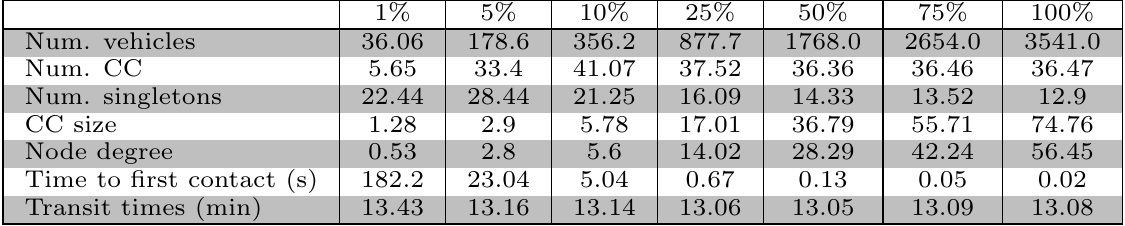}
  \label{table:dataset}
\end{table}

We define a contact as a robust communication that allows reliable data delivery between two vehicles. We assume that all the vehicles may communicate in an ad hoc fashion using the IEEE~802.11 amendment for Wireless Access in Vehicular Environments (WAVE)~\cite{80211p}. As wireless propagation models are not the core of this paper, we assume a deterministic model where a packet is successfully received if the receiver's distance is below a certain indicative value. Following a pragmatic approach, we consider path loss model approximations and measurements in an urban line-of-sight environment performed by Cheng \textit{et al.}~\cite{cheng2007mobile}, both corroborating on the existence of a critical distance at $d=100$~m, above which radio propagation suffers from high degradation and variability. Vehicles less than $100$~m apart were therefore considered within transmission range of each other (i.e., in ``contact''). Furthermore, we define the connectivity graph as a time-variant undirected graph with mobile nodes as vertices. Mobile nodes are connected if a contact exists between them.

However, we do not expect all vehicles in Bologna to participate in a single Push-and-Track process for a variety of reasons that range from market penetration rate, to incompatible manufacturer implementations, or simply different unrelated content feeds. In order to simulate this, we derive new datasets composed of only a certain percentage of the vehicles in the original dataset. For example a ``1\%'' dataset comprises a subset of the vehicles of the original dataset that were chosen with a uniform random 1\% probability. In practice, we derived 10 such datasets for each of the following percentages: 1\%, 5\%, 10\%, 25\%, 50\%, and 75\%. With the exception of the ``100\%'' case, all measurements presented in this section are averages over 10 derived datasets.

\begin{figure}
  \centering
  \includegraphics{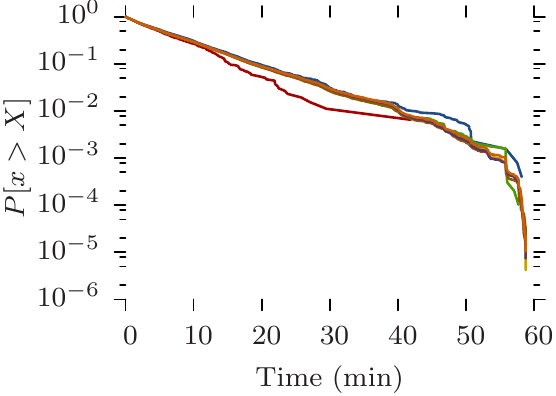}
  \caption{Complementary cumulative distribution (ccdf) of contact times for all percentages of participating vehicles.}
  \label{fig:ct_cdf}
\end{figure}

Some metrics are preserved through this transformation, while others vary significantly. Table~\ref{table:dataset} sums up the evolution of the relevant metrics for this work.

Among those that are preserved are the contact duration and transit times distributions. The distribution of contact duration follows an exponential distribution with the same parameter for all percentages (see Fig.~\ref{fig:ct_cdf}). Most contacts are short lived (50\% last less than 25 seconds), illustrating the highly dynamic nature of the vehicular mobility, but a few last up to 50 minutes. Similarly, transit times are identical for all participation percentages.

Other metrics, such as the average number of vehicles, the average size of connected components (i.e., the number of nodes in the component), and the average node degree, vary linearly with the participation rate.

\begin{figure}
  \centering
  \includegraphics{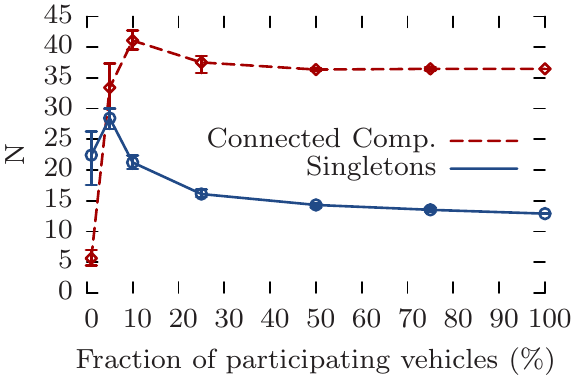}
  \caption{Average number of connected components and singletons for different percentages of participating vehicles. Each point is averaged over 10 runs; bars indicate the maximum and minimum value for those 10 runs.}
  \label{fig:num_cc_sing}
\end{figure}

The average number of connected components, depicted in Fig.~\ref{fig:num_cc_sing}, has an interesting property. When the number of participating nodes initially increases, the number of connected components sharply increases before reaching a maximum around 45 at 10\%. Beyond 25\% its value is stable at about 35.  Similarly the average number of singletons increases sharply before slowly declining to only 15. This leads to two observations. Firstly,  despite the important number of vehicles and the presence of some large connected components (up to 1,200 nodes), the network remains highly partitioned at all times with a large amount of isolated vehicles. In good opportunistic fashion, exploiting node mobility is therefore crucial to achieving connectivity over time. Secondly, combined with the linear increase of node degree and connected component size, this means that beyond 25\% of participation rate, all new nodes mostly densify \emph{existing} connected components (without extending the area covered by these), while the basic structure of the network remains the same. This will have an important impact on the \emph{periodic flooding} scenarios described in Section~\ref{subsec:periodic_scenario}.

Finally, the \emph{time to first contact}, i.e., the time between the moment a node subscribes to a content and the time it meets another node, decreases very quickly with the participation rate. It ranges from about 3 minutes at 1\% to nearly instantaneous at 25\% and higher. This suggests that the \emph{floating data} scenario, described in Section~\ref{subsec:floating_scenario}, will be challenging for low participation rates and easy at 25\% and above.

\section{Simulation setup}
\label{sec:setup}

\subsection{Simulator}
\label{subsec:simulator}

The results in the following sections are all based on the Bologna car traffic dataset from a typical weekday between 8~a.m. and 9~a.m. described in the previous section. Unfortunately, none of the existing network simulators we surveyed~\cite{ONE,ns3} were adapted to evaluate Push-and-Track strategies, not to mention severe scalability issues when simulating several thousand users. For the purposes of this paper, we built our own simulator, heavily inspired by the ONE DTN simulator~\cite{ONE}, and built on top of the DITL library~\cite{DITL}. In particular, it retains the contact-based ad hoc communication model from ONE, with its simple interference model in which a node may only communicate with a single neighbor at any given time. Unlike ONE, all routing is broadcast, there are different classes of messages (e.g., content or control), and different wireless media (e.g., infrastructure and ad hoc). While techniques such as fountain-codes~\cite{Luby2002} make reliable point-to-multipoint transmissions of large messages feasible at the link layer, in this paper only point-to-point communications are considered and reliability is provided at the transport layer (e.g. TCP). Therefore when a node wishes to transmit a copy of the content to each of its neighbors, it must iteratively transmit the full content to each one. Furthermore, we assume that each user has a non-interfering infrastructure link to the control system with different upload and download rates.

Vehicles send \textsf{ENTER}, \textsf{LEAVE}, and \textsf{ACK} control messages as described in Section~\ref{subsec:control-loop}. As for the optional messages, we set a timer of one minute for both the GPS-based and Connected Components strategies.

All transfers, including control messages, are simulated and may fail. An ad hoc transfer will fail if either the nodes move out of range of each other or one of the nodes leaves the area before the end of the transfer. An infrastructure transfer, with the exception of the \textsf{LEAVE} messages, will also fail if the node leaves the area too early. Furthermore, a node may be simultaneously receiving the same message from both the infrastructure and from another node via its ad hoc interface; whichever one finishes first cancels the other. The amount transferred before the cancel of course counts against the total loads for ad hoc or infrastructure.

\subsection{Parameters}
\label{subsec:exp_setup}

As in any simulation, the values of certain parameters inevitably incur some arbitrariness. We tried to keep this to a minimum. The bit-rate of the ad hoc links is set to 1~Mbytes/s which is compatible with the IEEE~802.11 amendment for wireless in vehicular environments (WAVE)~\cite{80211p}. The bit-rate for the infrastructure downlink is set to 100~Kbytes/s. This is twice the expected bit-rate of EDGE networks but much less than the advertised 7.2~Mbits/s rate of HSDPA. However, surveys in Europe and the US have shown that the average user-experienced 3G downlink rate is typically just below 128~Kbytes/s~\cite{wired_3g,ufc_3g}. The infrastructure uplink rate is set to 10~Kbytes/s. Furthermore, each content message is set to 1~Mbyte in size. This means that it takes 10 seconds to transfer over the infrastructure and 1 second over the ad hoc link. The bit-rates that we consider here might either be optimistic or pessimistic depending on nodes location, velocity, or on the access networks they use. Because our evaluation is meant to demonstrate how Push-and-Track can leverage opportunistic communications, we make simplistic assumptions on low layers, and leave more accurate evaluations for future work. Finally, for the sake of simplicity, control messages are all 256-bytes long. This is probably excessive for simple \textsf{ENTER}, \textsf{LEAVE}, and \textsf{ACK} messages, but long enough to accommodate a sizable list of neighbors.

\subsection{Metrics}
\label{subsec:metrics}
This paper is focused on offloading traffic from the infrastructure to the ad hoc radio. Our analysis will focus on both aggregate and dynamic metrics.

The aggregate metrics are the \emph{infrastructure load}, the \emph{ad hoc load}, and the \emph{offload ratio}. The infrastructure (resp. ad hoc) load is the sum (in bytes) of all transfers, successful, failed, or aborted, using the infrastructure (resp. ad hoc) radio during a 1-hour simulation run. The load induced by control messages is of course included in the total infrastructure load but is typically one or more orders of magnitude less than the load incurred by pushing the content to nodes. The offload ratio is computed by comparing the infrastructure load of a specific run with the reference infrastructure load \emph{in the absence of any ad hoc radio}. Formally, let $L$ be the infrastructure load of a simulation run, and $L_{ref}$ be the load when only using the infrastructure, then the offload ratio is $1-L/L_{ref}$.

The dynamic metric is the \emph{infection ratio}. For a given message at time $t$, the infection ratio is the fraction of nodes that have received the message. Although mostly increasing, the infection ratio may decrease when new nodes enter the simulation area.

\section{Results 1: Periodic flooding}
\label{sec:periodic}

In this section, we present offloading results for Push-and-Track in the periodic flooding scenario defined in Section~\ref{subsec:periodic_scenario}.

\subsection{Preliminaries}
\label{subsec:periodic_preliminaries}

Messages are sent periodically, with the previous one expiring as the next one is sent. Even though our simulator can handle multiple competing messages, in order to properly identify the important factors influencing message propagation, we limited ourselves to a single message at any given time in the network. In this paper, two message lifetime periods were tested: a tight 1-minute delay and a more relaxed 10-minute delay. As we will see, the results differ significantly between these two constraints.

Benchmarks for both delay constraints were established using the \textit{Infrastructure only} and \textit{Dominating set oracle} strategies. Furthermore, each pair of \textit{when} and \textit{whom} strategies described in Section~\ref{subsec:periodic_scenario} was tested. A run spans the full hour of the dataset and consists in periodically sending a new message and then controlling its propagation using a particular strategy pair. Therefore, in the 1-minute scenario, 60 messages will be sent per run, as opposed to just 6 in the 10-minute scenario.

The latency of the infrastructure links (10 seconds in our example) imposes a delay between the moment when a re-injection decision is taken, and the moment when that decision has an effect on the epidemic propagation. This is particularly tricky during the first 10 seconds when no copies have yet begun disseminating in the ad-hoc network. During that time, the feedback loop is essentially blind. To allow for pushed messages to begin propagating and have an impact before triggering a new re-injection decision, we set $\Delta_t$, the update time of the control loop, to 20 seconds, i.e., twice the transmission time in our example. Furthermore, to prevent all strategies, fast or slow, from initially sending a single copy right after a new message is created by the controller, the first injection decision occurs 1 second later. For example, after that 1 second, a \textit{Square Root} strategy will inject more initial copies than a \textit{Quadratic} one.

This section presents two types of results: global averages (Figs.~\ref{fig:infra_relief}, \ref{fig:matrix_1}, and~\ref{fig:matrix_10}) and dynamic averages (Figs.~\ref{fig:dynamic_1} and~\ref{fig:dynamic_10}). The global results are averages over 10 runs, i.e. over 600 messages in the 1-minute scenario, and 60 messages in the 10-minute scenario. In order to smooth out effects due to the particular network topology at the beginning of each period, the sending time of the first message is shifted by $T/10$ at every subsequent run, where $T$ is the sending period (i.e., the message lifetime). The dynamic results are also averages over 10 runs but are focused on a specific period and hence without any shifting of the sending time of the first message.

\begin{figure}[t]
  \centering
  \includegraphics{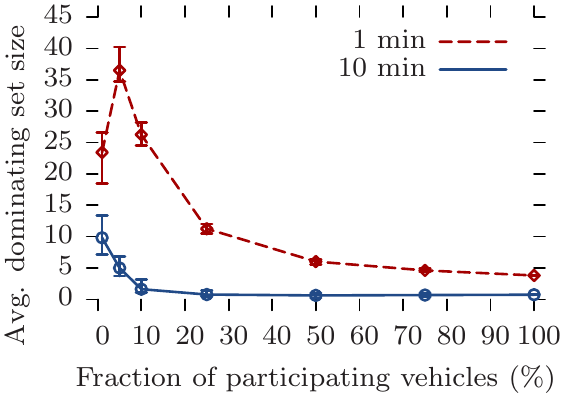}
  \caption{Average number of vehicles in the dominating set for different participation rates. Results show for maximum message delays of 1 and 10 minutes. Each point is averaged over 10 runs; bars indicate the maximum and minimum value for those 10 runs.}
  \label{fig:ds_size}
\end{figure}

Finally, results using the dominating set oracle, which precalculates at each new message's creation time a small set of nodes with space-time paths to all other nodes before that message's expiration time (see Section~\ref{subsubsec:benchmarking}), are provided for comparison. Fig.~\ref{fig:ds_size} shows the average number of nodes in the dominating set for different participation rates and maximum message delays. In the 1-minute maximum delay case, this dominating set can comprise up to 35 nodes with a 10\% participation rate, before declining to just 5 nodes when all nodes participate. With the longer 10-minute delay the dominating set is usually reduced to just one node. Keep in mind that this oracle ignores ad hoc transmission times. As we shall see, this has a big impact on its performance in the short 1-minute delay runs.

\subsection{Relieving the infrastructure}
\label{subsec:relief}

\begin{figure}[t]
  \centering
  \subfloat[1 min delay\label{subfig:infra_relief_1}]{\includegraphics{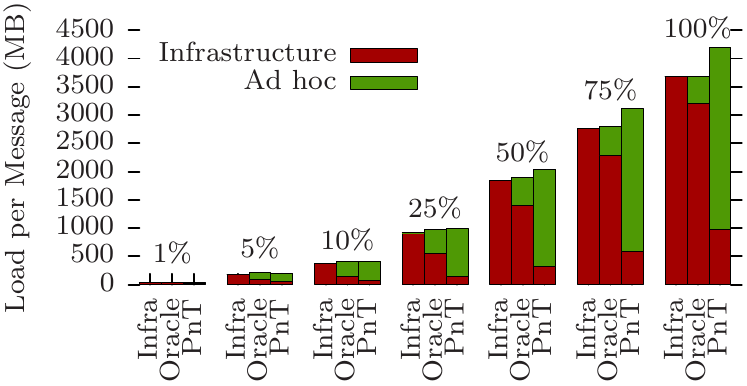}} \\
  \subfloat[10 min delay\label{subfig:infra_relief_10}]{\includegraphics{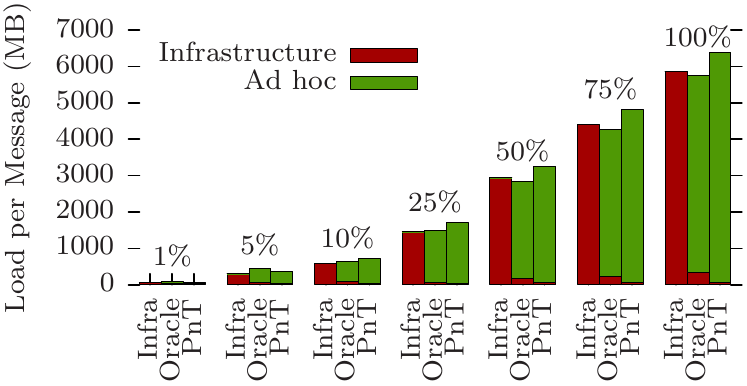}}
  \caption{Infrastructure vs. ad hoc load per message sent using only the infrastructure (Infra), Push-and-Track (PnT), and the Dominating Set Oracle (Oracle).}
  \label{fig:infra_relief}
\end{figure}

Push-and-Track does an excellent job of relieving the load on the infrastructure by transferring most of it to faster and cheaper ad hoc communications. Fig.~\ref{fig:infra_relief} shows the average total amount of information transferred per message and how this is split between infrastructure and ad hoc. The results for Push-and-Track in this figure correspond to the best \textit{when} and \textit{whom} pair of strategies for a 10-minute or 1-minute delay. How the different strategies combine and compare to each other will be examined in the following sections. In Fig.~\ref{fig:infra_relief}, the totals for a 10-minute delay are greater than those for a 1-minute delay. Indeed, since most vehicle transit times are less than 10 minutes (see Section~\ref{sec:dataset}), there are more vehicles in the simulation area over a 10-minute period than a 1-minute period, hence the difference in total transfer amounts per message.

Push-and-Track manages to transfer nearly all of the load from the infrastructure to ad hoc communications: 98\% for a 10-minute delay, and up to 85\% for a 1-minute delay (for the 25\% participation rate). This is the main result of this paper: \emph{thanks to its feedback loop, the Push-and-Track framework can achieve offload ratios around 90\% while guaranteeing 100\% on time message delivery}. The ratio is less good with a tighter delay simply because the epidemic ad hoc dissemination has less time to propagate the message to the entire network and thus more copies must be re-injected to parts of the network that have not yet received the content. 

With a 1-minute delay, the Push-and-Track offload ratio initially increases with the participation rate, reaches a maximum at 25\%, and then degrades beyond that (Fig.~\ref{subfig:infra_relief_1}), whereas with a 10-minute delay, it increases monotonously and caps out at 25\%. In Section~\ref{subsec:dataset_analysis}, we showed how the datasets connected-components structure was achieved with a 25\% participation rate. All new nodes beyond that mostly densify existing connected components. In a sense, the ad hoc offloading can already be fully harnessed when the participation rate is 25\%. This explains why the performance in the 10-minute scenarios cap out at 25\%. However, since the ad hoc epidemic dissemination uses point-to-point communications at each hop, denser connected components also slow down its geographic propagation.  With a 10-minute delay this is not a problem, but under a tight 1-minute delay, a message does not have time to completely disseminate through a large connected component, thereby requiring more re-injections and degrading the offload ratio.

Interestingly, the best Push-and-Track strategy pair almost always beats the \textit{Dominating Set Oracle} in terms of offload ratio. This is particularly visible with a 1-minute delay. In this case, the oracle's offload ratio actually degrades for participation rates above 10\%. At 100\%, its offload ratio is a measly 12\%. As previously, this is mainly due to the epidemic propagation not having time to fully explore every space-time path within 1 minute. For example, if a node from a large connected components moves to another large connected component late during the 1-minute period, the oracle will assume there exists a space-time path from any node in the first connected component to any node in the second one. However this does not mean that injecting one copy into the first connected component will guarantee that everyone in the second connected component will be infected before the end of the message's lifetime. Because of this, the oracle hits the ``panic zone'' (see Section~\ref{subsec:periodic_scenario}) before having infected every node. Whatever efficiency is gained by an excellent choice of initial nodes to infect is lost when it has to push the content to all remaining uninfected nodes as the deadline gets close. On the other hand, Push-and-Track, by keeping track of the epidemic's progression and re-injecting copies when needed, is less affected by the ``panic zone'' and thus can outperform the oracle despite making poorer choices of whom to push to. This underscores one of our main arguments in this paper: \textit{having a feedback loop for re-injecting content is essential for guaranteeing delivery delays in a hybrid infrastructure/ad hoc network}.

\subsection{Random is difficult to beat}
\label{subsec:beating_random}

When surveying the results for all \textit{when} and \textit{whom} strategy pairs, the \textit{Random} re-injection strategy consistently does better than most of the more sophisticated strategies described in Section~\ref{sec:strategies}. This section examines this observation in more detail and studies the impact of \textit{whom}-strategies on the infrastructure load.

\begin{figure}[t]
  \centering
  \scalebox{0.85}{
  \subfloat[1\%\label{subfig:matrix_p001_1}]{
    \includegraphics{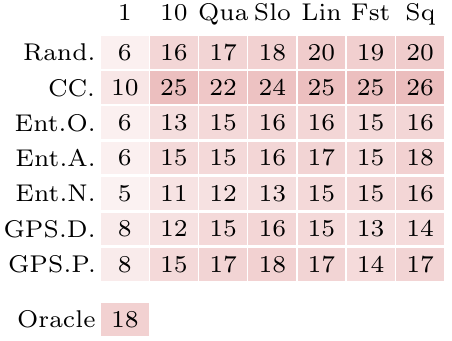}}
  \subfloat[25\%\label{subfig:matrix_p025_1}]{
    \includegraphics{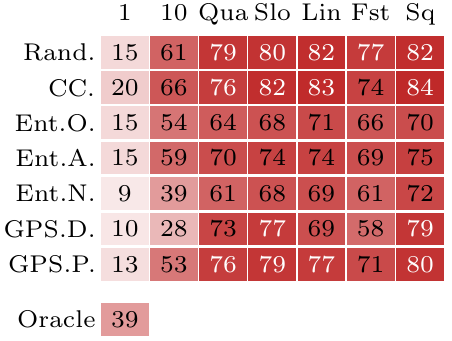}}
  \subfloat[100\%\label{subfig:matrix_p1_1}]{
    \includegraphics{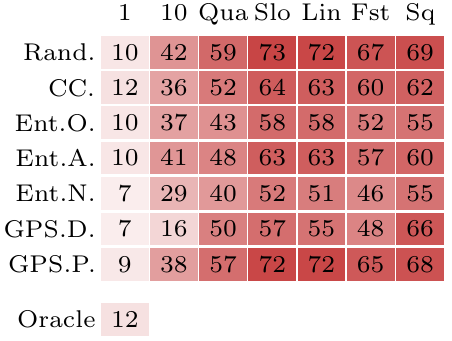}}
  }
  \caption{1-minute delay: average offload ratio for different combinations of \textit{whom} and \textit{when} strategies. The rows correspond, from top to bottom, to the following \textit{whom} strategies: Random, Connected Components, Entry-Oldest, Entry-Average, Entry-Newest, GPS-Density, and GPS-Potential. The columns represent the following \textit{when} strategies, from left to right, Single Copy, Ten Copies, Quadratic, Slow Linear, Linear, Fast Linear, and Square Root.}
  \label{fig:matrix_1}
\end{figure}

\begin{figure}[t]
  \centering
  \scalebox{0.85}{
  \subfloat[1\%\label{subfig:matrix_p001_10}]{
    \includegraphics{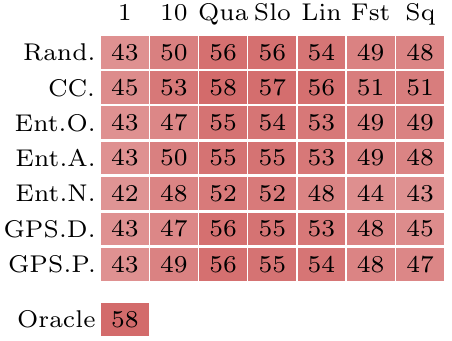}}
  \subfloat[25\%\label{subfig:matrix_p025_10}]{
    \includegraphics{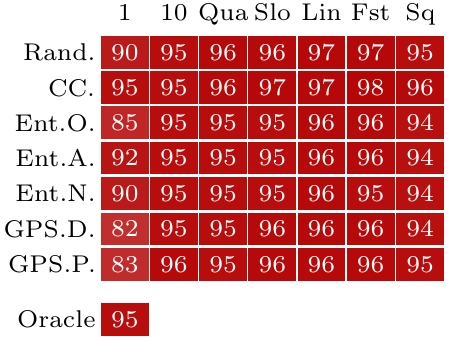}}
  \subfloat[100\%\label{subfig:matrix_p1_10}]{
    \includegraphics{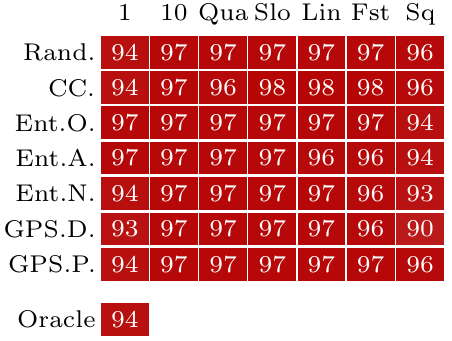}}
  }
  \caption{10-minute delay: average offload ratio for different combinations of \textit{whom} and \textit{when} strategies. The rows correspond, from top to bottom, to the following \textit{whom} strategies: Random, Connected Components, Entry-Oldest, Entry-Average, Entry-Newest, GPS-Density, and GPS-Potential. The columns represent the following \textit{when} strategies, from left to right, Single Copy, Ten Copies, Quadratic, Slow Linear, Linear, Fast Linear, and Square Root.}
  \label{fig:matrix_10}
\end{figure}

Figs.~\ref{fig:matrix_1} and~\ref{fig:matrix_10} plot the offload ratio for all combinations of \textit{whom} and \textit{when} strategies for, respectively, a 1-minute and a 10-minute delay. Each square in these matrices is colored from white (0\%) to red (100\%). The infrastructure load measurements naturally include the control load. With a 10-minute delay, this amounts to roughly 3~Mbytes per message, except for the \textit{GPS-based} and Connected Components (\textit{CC}) strategies, where it goes up to 15~Mbytes per message due to the periodic updates on current position or current neighbors. With a 1-minute delay, those numbers become 1~Mbyte and 2~Mbytes, respectively. In any case, they remain small compared to the total load on the downlink.

With a 10-minute delay, a few initial copies have enough time to propagate epidemically to all the vehicles in the simulation area. In fact, beyond a 25\% participation rate, a single well placed copy is enough to basically reach all the vehicles (Fig.~\ref{subfig:matrix_p025_10}). At 100\% participation rate that single copy does not even need to be very well placed, although if a choice must be made, prioritizing nodes that have been in the simulation area for a while (i.e., the \textit{Entry-Oldest} or \textit{Entry-Average} strategies) provides an edge. However, regardless of the participation rate, the best strategy never improves on \textit{Random} by more than a few percentage points. These results suggest that, with a 10-minute delay, advanced re-injection strategies are not really required.

With a 1-minute delay, nearly every strategy performs significantly worse than \textit{Random}. In particular, the \textit{GPS-Density} strategy frequently targets nodes that are both in the same dense connected component, leading to many ``useless'' pushes. The \textit{GPS-Potential} improves on this by spreading the copies to the least infected areas, but, because of this, will frequently push to nodes in areas of very sparse connectivity. The \textit{Entry-Newest} and \textit{Entry-Oldest} tend to target nodes on the edge of the simulation area, whereas the \textit{Entry-Average} targets node closer to the center. \textit{Random} combines the best of all these strategies. Indeed it statistically has a high chance of hitting the large connected components and also tends to spread the copies uniformly over the area. Again, the only strategy that beats it is the \textit{CC} strategy for percentage ratios below 25\%. Starting at 50\%, the \textit{Random} strategy overtakes the \textit{CC} strategy. Indeed, the \textit{CC} strategy targets connected components that do not yet have an infected node in it (see Section~\ref{subsec:periodic_scenario}). As previously, one minute is not enough for the message to propagate through a large connected component from only a single initially infected node. On the other hand, the larger a connected component, the more likely the \textit{Random} strategy will push several copies to it, thereby increasing the likelihood that the epidemic dissemination will cover it before the 1-minute delay ends.

If one is not willing to deal with the added complexity of a more sophisticated control channel, let alone privacy concerns about localization and/or proximity information, then the simple \textit{Random} \textit{whom}-strategy consistently performs very well. While this conclusion is specific to our vehicular scenario, we believe that it holds for any scenario in a limited geographic area with many large connected components.

\subsection{Fast or slow start?}
\label{subsec:fast_slow}

We examine how the infection ratio evolves over the course of one message's lifetime for different \textit{when-strategies}. All results in this section use the \textit{Random} \textit{whom-strategy}. What is the better strategy: sending many initial copies, in order to avoid the ``panic zone'', or few, at the risk of having to push extra copies as the deadline gets close? 

\begin{figure}[t]
  \centering
  \subfloat[1\%\label{subfig:p001_fast_1}]{
    \scalebox{0.8}{\includegraphics{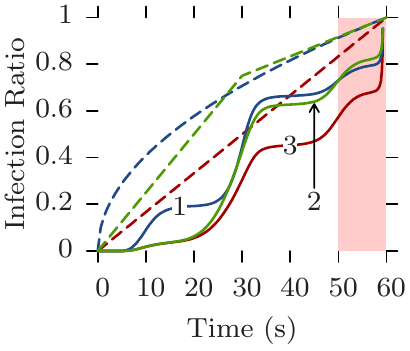}}} \quad
  \subfloat[25\%\label{subfig:p025_fast_1}]{
    \scalebox{0.8}{\includegraphics{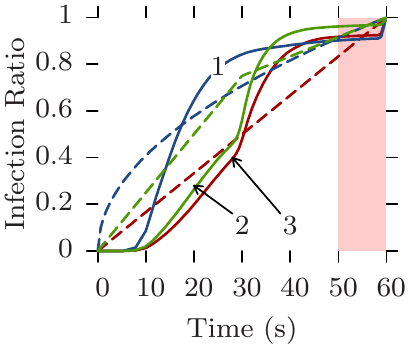}}} \quad
  \subfloat[100\%\label{subfig:p1_fast_1}]{
    \scalebox{0.8}{\includegraphics{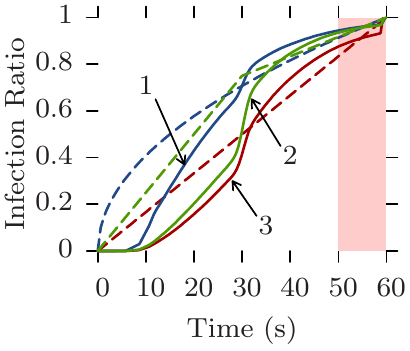}}} \\
  \subfloat[1\%\label{subfig:p001_slow_1}]{
    \scalebox{0.8}{\includegraphics{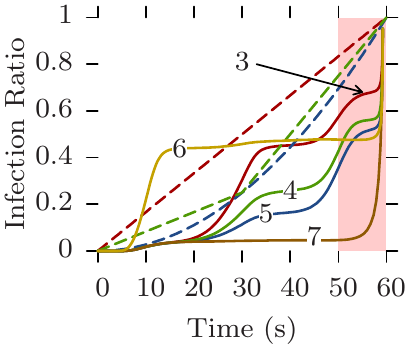}}} \quad
  \subfloat[25\%\label{subfig:p025_slow_1}]{
    \scalebox{0.8}{\includegraphics{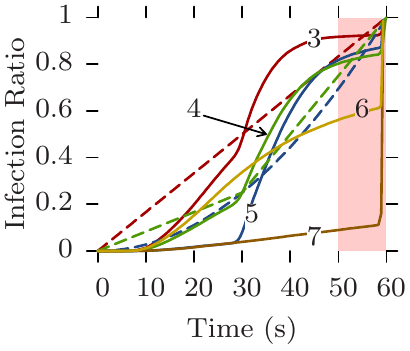}}} \quad
  \subfloat[100\%\label{subfig:p1_slow_1}]{
    \scalebox{0.8}{\includegraphics{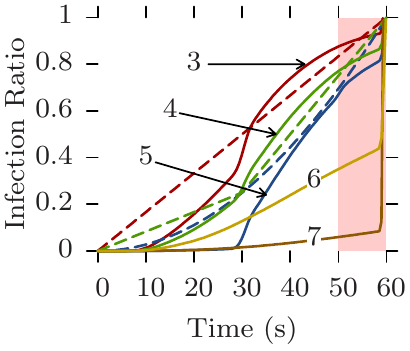}}}
  \caption{Infection rates with 1-minute maximum delay depending the when-strategy. All results are for the \textit{Random} re-injection strategy. Objective functions are dashed and the light red area corresponds to the ``panic zone''. The fast start strategies -- (1) Square root, (2) Fast linear, (3) Linear -- are on the top row, while the slow start ones -- (3) Linear, (4) Slow linear, (5) Quadratic, (6) Ten copies, (7) Single copy -- are on the bottom row.}
  \label{fig:dynamic_1}
\end{figure}

Fig.~\ref{fig:dynamic_1} shows the evolution of the infection ratio for various slow-start and fast-start strategies with a 1-minute delay. The corresponding objective functions are represented by dashed lines and the panic zone is the light red area. On all sub-plots, the infection ratio is zero for the first ten seconds, which is the time required to send the first copies over the infrastructure. However, from the point of view of the control system, a node is considered infected as soon as a transfer is initiated to avoid any explosion in the number of initiated transfers. The latency of the infrastructure links (10 seconds in our example) imposes a delay between the moment when a re-injection decision is taken, and the moment when that decision has an effect on the epidemic propagation.

Figs.~\ref{subfig:p001_fast_1} and~\ref{subfig:p001_slow_1} plot the infection ratios with a 1\% participation rate. As stated before, in this case, the network is highly disconnected (e.g., the average node degree is less than 1, see Section~\ref{subsec:dataset_analysis}). The epidemic dissemination does not really help. For example, the infection ratio of the \textit{Ten Copies} strategy on Fig.~\ref{subfig:p001_slow_1} remains roughly constant, forcing the controller to re-inject many copies upon entering the ``panic zone''.  With such a low participation rate, the infection ratio is constantly behind the objective function. Therefore, every 20 seconds, the controller re-injects a large number of copies, leading to a ``step'' pattern for both fast and slow strategies.

This ``step'' pattern progressively disappears for higher participation rates. On the 25\% plots (Figs.~\ref{subfig:p025_fast_1} and~\ref{subfig:p025_slow_1}), sudden slope changes are still visible around the 30 seconds mark. These are the result of the re-injection at 20 seconds. Once the epidemic propagation kicks in however, the infection ratio grows rapidly and eventually overtakes the objective function. Surprisingly, this behavior is \emph{less} pronounced when the participation rate is 100\%. Here again, the increased density of participants slows the epidemic propagation. This accounts for the lower overall offload ratio with a 100\% participation as opposed to just 25\% (Fig.~\ref{subfig:matrix_p025_1}).

Every push decision by the infrastructure has an associated cost/benefit analysis. The cost is of course incurred by using the infrastructure. The benefit is the cost of spared pushes (i.e., the number of copies it will contribute to final number of infected nodes) minus the cost of the push. This benefit can be positive, and, due to the epidemic propagation, the sooner the copy is pushed, the more likely it is to be positive. It can be nil, if the copy is sent to an isolated node, or during the ``panic zone'' where it will not have time to replicate to others after being received. Finally the benefit can be \emph{negative}. For example, a copy pushed to a node that would have received the message shortly thereafter through the epidemic propagation will likely contribute no new infections, and is thus wasted.

In the absence of a control loop, the choice of the initial number of copies to inject has a huge impact on the offload ratio. Consider for example the \textit{Single Copy} and \textit{Ten Copies} strategies with a 25\% participation rate and 1-minute delay on  Fig.~\ref{subfig:p025_slow_1}. Both of these just send an initial number of copies and then wait until the panic zone. Due to the epidemic propagation, a difference of 9 in the initial number of copies translates on average to 400 extra copies right before the ``panic zone''. Intuitively, it should be possible to estimate, on a given scenario, the ``optimal'' initial number of copies to initially inject to cover nearly everyone before triggering the ``panic zone''. However such a strategy would only be optimal \emph{on average}, as it would not be able to react to changes in the number of vehicles or dramatic network splits. In addition, it would also require a training period. 

On the other hand, the control loop means that Push-and-Track can quickly react and adapt to changing conditions. For example, even though the \textit{Slow Linear}'s infection ratio is initially below that of the \textit{Ten Copies} strategy (Fig.~\ref{subfig:p025_slow_1}), the \textit{Slow Linear} strategy reacts to the slow infection ratio by re-injecting new copies after 20 seconds. This allows it to avoid the massive last-minute re-injections upon arriving in the panic zone, and achieve excellent offloading performance (82\% vs 61\%, Fig.~\ref{subfig:matrix_p025_1}).

The \textit{Fast Linear} strategy is interesting in that its average performance, regardless of which \textit{whom}-strategy it is associated with, tends to be inferior to both the \textit{Square Root} and \textit{Linear} strategies (Figs.~\ref{subfig:matrix_p001_1}, ~\ref{subfig:matrix_p025_1}, and~\ref{subfig:matrix_p1_1}). The dynamics of its infection ratio for a 25\% participation rate (Fig.~\ref{subfig:p025_fast_1}) shows why. The \textit{Square Root} strategy initially overshoots its objective but then does not need to re-inject copies until the very end. The \textit{Linear} strategy fails to meet its objective after 20 seconds, and thus begins to re-inject copies. Similarly, the \textit{Fast Linear} strategy also fails to meet its objective after 20 seconds, but it fails by a much greater margin. It then over-reacts and re-injects too many copies. \emph{Therefore fast vs slow start is perhaps an oversimplification of the problem. With a 1-minute delay, how the objective function and the induced epidemic dissemination interact is what matters most.}

\begin{figure}[t]
  \centering
  \subfloat[1\%\label{subfig:p001_fast_10}]{
    \scalebox{0.8}{\includegraphics{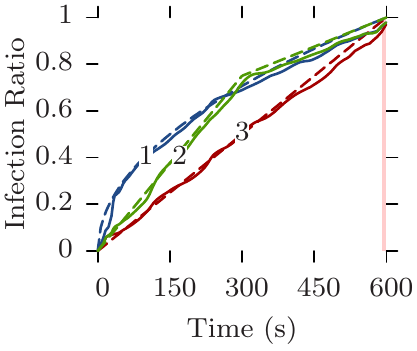}}} \quad
  \subfloat[25\%\label{subfig:p025_fast_10}]{
    \scalebox{0.8}{\includegraphics{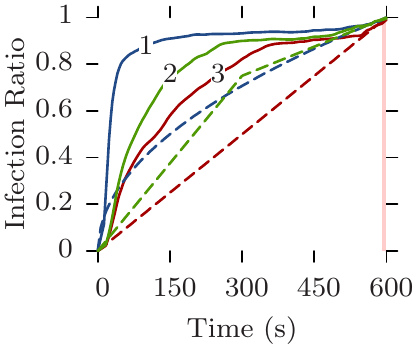}}} \quad
  \subfloat[100\%\label{subfig:p1_fast_10}]{
    \scalebox{0.8}{\includegraphics{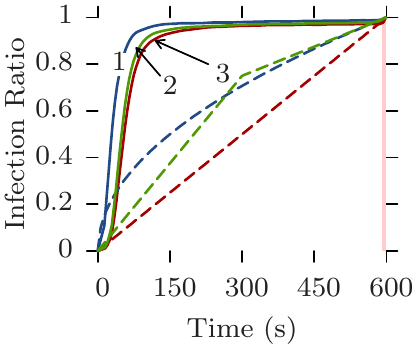}}} \\
  \subfloat[1\%\label{subfig:p001_slow_10}]{
    \scalebox{0.8}{\includegraphics{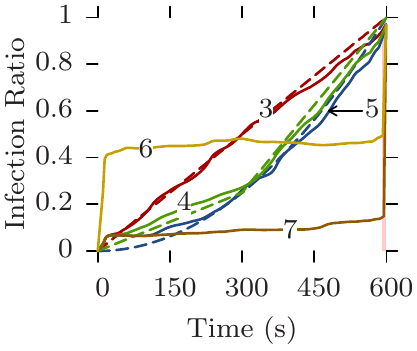}}} \quad
  \subfloat[25\%\label{subfig:p025_slow_10}]{
    \scalebox{0.8}{\includegraphics{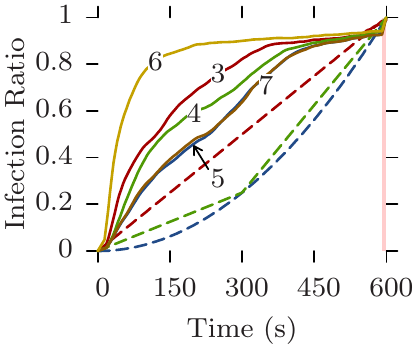}}} \quad
  \subfloat[100\%\label{subfig:p1_slow_10}]{
    \scalebox{0.8}{\includegraphics{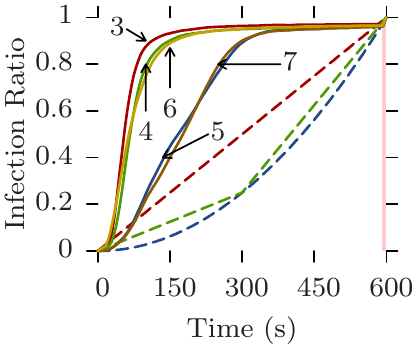}}}
  \caption{Infection rates with 10-minute maximum delay depending the when-strategy. All results are for the \textit{Random} re-injection strategy. Objective functions are dashed and the light red area corresponds to the ``panic zone''. The fast start strategies -- (1) Square root, (2) Fast linear, (3) Linear -- are on the top row, while the slow start ones -- (3) Linear, (4) Slow linear, (5) Quadratic, (6) Ten copies, (7) Single copy -- are on the bottom row.}
  \label{fig:dynamic_10}
\end{figure}

Fig.~\ref{fig:dynamic_10} plots the same metrics as Fig.~\ref{fig:dynamic_1} except with a 10-minute delay. Since the control loop still makes re-injection decisions ever 20 seconds, a longer delay smooths out the ``step'' pattern seen with a 1-minute delay. For example, with a 1\% participation rate, all strategies closely follow their objective functions (Fig.~\ref{subfig:p001_fast_10}), instead of the ``step'' patterns with a tighter 1-minute delay (Fig.~\ref{subfig:p001_fast_1}). Even though the network is very sparse with only 1\% of participating vehicles, having the control loop regularly injecting new copies to match an objective function does improve the average offload rate by approximately 10\% on average (Fig.~\ref{subfig:matrix_p001_10}).

With higher participation rates (e.g., 25\% and 100\%), for a 10-minute delay, all strategies rapidly overtake their objective functions. For example, with a 100\% participation rate, all slow-start strategies achieve near-complete coverage after only 5 minutes (Fig.~\ref{subfig:p1_fast_10}). This means that when long delays are acceptable, then all strategies are roughly equivalent. In this case, they differ mainly in the initial number of pushed copies. However, particularly when the number of vehicles is several hundreds or thousands, the offload ratio is impacted more by the singletons in the panic zone than by the number of initial copies. This explains why all pairs of \textit{whom} and \textit{when} strategies lead to equivalent offload ratios with a long 10-minute delay (Fig.~\ref{fig:matrix_10}). Hence, \emph{the feedback loop and periodic re-injections are crucial
under tight delay constraints, but less so under more relaxed ones.}

\section{Results 2: Floating data}
\label{sec:floating}

In this section we present offloading results for Push-and-Track in the floating data scenario defined in Section~\ref{subsec:floating_scenario}. A unique content message is created for the entire 1-hour run and is replicated among the participating vehicles. The focus here is to guarantee that \emph{new} vehicles entering the simulation area will receive the messages within a certain maximal delay after the moment they subscribe.

In order to measure how Push-and-Track can bring delivery and delay guarantees to the floating data scenario, one must first understand how the system behaves without a feedback loop. For each participation rate (1\%, 5\%, 10\%, 25\%, 50\%, 75\%, and 100\%), we started from a state in which all vehicles initially present in the simulator have a copy of the content. We then track, for all new vehicles if they receive the content before leaving and, if so, how long they had to wait for it.

\begin{figure}[t]
  \centering
  \includegraphics{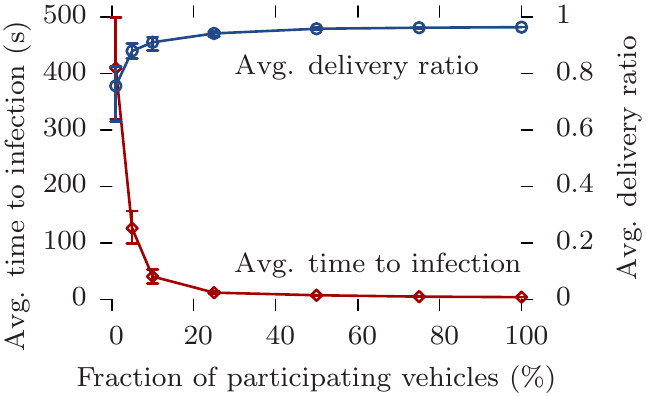}
  \caption{Floating data performance without any feedback loop for different participation rates. Each point is averaged over 10 runs; bars indicate the maximum and minimum value for those 10 runs.}
  \label{fig:floating_nopnt}
\end{figure}

Fig.~\ref{fig:floating_nopnt} plots the average delivery ratio and average time to infection (i.e., the content delivery delay), for all participation rates. Average delivery ratio starts at about 0.8 when only 1\% of vehicles participate and maxes out at 0.95 when all vehicles participate. In parallel, the average time to infection sharply drops from over 5 minutes at 1\% to a couple of seconds at 25\% and beyond. Furthermore, the time to infection shows strong variations for low participation rates. This analysis shows that there exists a strong case for Push-and-Track in the floating data scenario. Indeed, in all cases, approximately 5\% of vehicles never receive the content, and, for low participation rates, delays are large and difficult to control.

Here we define the maximum floating delay as the amount of time allowed for a node to receive the content via the ad hoc radio. If it has not received the content within that time, then the infrastructure will push the content to it. Note that a maximum floating delay of $0$ is different from the infrastructure-only scenario. In both cases, the infrastructure immediately starts pushing the content to new subscribers. However in the former case, nodes may also receive the content via the ad hoc radio and then abort the infrastructure transfer.

\begin{figure}[t]
  \centering
  \includegraphics{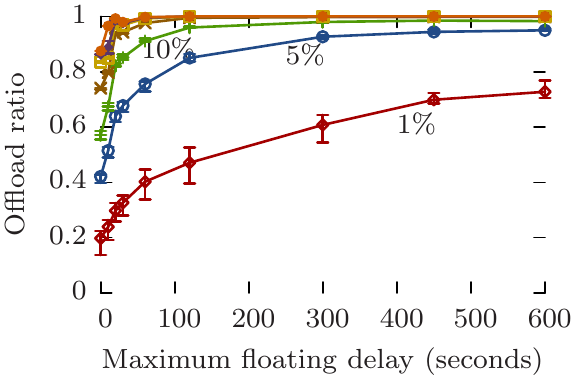}
  \caption{Floating data offload ratio vs. the maximum floating delay. For participation rates above 25\%, all plots are basically identical. Each point is averaged over 10 runs; bars indicate the maximum and minimum value for those 10 runs.}
  \label{fig:floating}
\end{figure}

Fig.~\ref{fig:floating} plots the Push-and-Track offload ratio results against the maximum floating delay for all participation rates. With a 1\% participation rate and tight maximum floating delay constraints, the infrastructure must support up to 80\% of the load. As the maximum floating delay is relaxed, the ad hoc radio progressively offloads more traffic, up to 72\% with a 10-minute maximum floating delay. For higher participation rates, the offloading ratios increase rapidly. For example, at 10\%, Push-and-Track guarantees a 20-second maximum floating delay while offloading 80\% of the traffic from the infrastructure to the ad hoc radio. With participation rates over 25\%, the offload ratio when waiting 10 minutes is over 99\%. \emph{Using a straightforward strategy, Push-and-Track can guarantee floating data performance while keeping infrastructure demand low.}

\section{Related work}
\label{sec:related}

Reducing the load on the wireless infrastructure has received attention in both academic and industrial circles. For example, Balasubramanian \textit{et al.} exploit the delay-tolerance of common types of data such as emails or file transfers to opportunistically offload them to available open WiFi hotspots~\cite{balasubramaniam:augmenting}. Similarly, Lee \textit{et al.} use a smartphone app to measure the availability of WiFi hot-spots in central Seoul for 3G offloading~\cite{Lee2010}. The now defunct French MVNO Ten Mobile had been offering free pushes of podcasts to their customers' mobile phone during the night using cheaper minutes~\cite{ten_mobile}. Every morning, users had the latest episodes of their favorite series pre-fetched on their mobile phones. More generally, opportunistic or delay-tolerant networks can exploit user mobility to increase an ad hoc network's capacity~\cite{GrossglauserTse02}. However, uncertain delays and probabilistic delivery ratios make such approaches unsuitable for most applications.

Cooperation between the wireless infrastructure and opportunistic networks is a hot topic that has begun to receive attention in the past couple of years. Hui et al. examine how hybrid infrastructure-opportunistic networks can improve delivery ratios over using either paradigm alone. In particular, they show that even infrastructure networks with high access point density can still significantly benefit from the opportunistic capabilities of its users~\cite{hui2009empirical}. More proactively, Ristanovic \textit{et al.} develop operator strategies for triggering 3G offloading to either infrastructure or ad hoc WiFi in certain areas in order to minimize smartphone energy usage~\cite{Ristanovic2011}. Using the cellular infrastructure as a control channel was first suggested by Oliver who exploits the low-cost of SMS to send small messages between participants in an opportunistic mobile network~\cite{oliver2008exploiting}. This idea was further expanded upon by Dimatteo \textit{et al.} through the MADNet architecture which integrates cellular, WiFi access point, and mobile-to-mobile communications for offloading traffic from the cellular network~\cite{Dimatteo2011}.

Closer to our work, Ioannidis \textit{et al.} push updates of dynamic content from the infrastructure to subscribers that then replicate it epidemically~\cite{ioa09}. The authors assume that the infrastructure has a maximum rate that it must divide among the subscribers. They then calculate the optimal rate allocation for each user in order to maximize the average freshness of content among all subscribers. Han et al. investigate different strategies to find the subset of the participants in a Mobile Social Network (MoSoNets) that will lead to the greatest infection ratio by the end of a message's lifetime. Therefore, pushing the content trough the cellular infrastructure to that optimal subset minimizes the load on the infrastructure~\cite{hui_offloading,Han2011}. Barbera \textit{et al.} analyze social encounter graphs according to various metrics to elect socially-important ``VIP delegates'', through whom 3G traffic will be offloaded to the other users~\cite{Barbera2011}. These papers are close to ours but differ in the following ways. Firstly, they do not have a feedback loop and cannot quickly react to changes in network dynamics or the arrival of new nodes. Secondly, the methods developed in both papers assume preexisting knowledge of pairwise contact probabilities. 

In a similar vein, Li \textit{et al.} study an opportunistic offloading scheme in which multiple contents are first pushed to special ``helper'' nodes that relay content opportunistically to subscriber nodes. Under the assumption of Poisson contact rates, they formulate the optimal offloading strategy as a submodular function maximization problem under multiple linear constraints
~\cite{Li2011}. In contrast, the Push-and-Track framework makes no assumptions on node mobility (the Poisson hypothesis for contact rate does not in fact hold for our dataset) and relies on full epidemic dissemination rather than only 2-hop helper-to-subscriber content propagation.

Finally, theoretical frameworks for determining liveness or expected lifetime of floating data have been developed in the context of sensors networks~\cite{Chakrabarti2007} and opportunistic networks~\cite{Hyytia2011}. In a vehicular scenario, hybrid infrastructure/opportunistic networks have been proposed for tying floating data to a given geographic area~\cite{Leontiadis2009,Jerbi2008}. However, unlike Push-and-Track, neither of these include a feedback loop and delivery remains probabilistic.

\section{Conclusions}
\label{sec:conclusion}

Push-and-Track is a framework for massively disseminating content with guaranteed delays to mobile users while minimizing the load on the wireless infrastructure. It leverages ad hoc communication opportunities, tracks the content spread through user-sent acknowledgments, and, if necessary, re-injects copies to nodes that have not yet received the content. Tests on the large-scale Bologna vehicular dataset reveal that Push-and-Track manages to reduce the infrastructure load by about 90\% while achieving 100\% delivery when periodically flooding content to a large number of subscribers, even under tight delay constraints. In this scenario, pushing content to random nodes works well as it manages to both hit the large connectivity clusters with high probability and spread the pushes uniformly around the city. Finally, Push-and-Track can also guarantee short delivery delays for floating data while keeping the infrastructure load very low. Push-and-Track's owes its adaptability and reactivity to its feedback loop.

Our work will continue in the following directions. Firstly, the feedback loop could be enhanced with a predictive epidemic propagation model. It could also take into account propagation measurement of previous messages to adjust its strategy for subsequent ones. Secondly, the impact of intermittent infrastructure connectivity must also be explored. Furthermore, this paper dealt with the case where all users were interested in the same content. However, the Push-and-Track framework is flexible and can be extended to a more realistic setting in which overlapping subsets of users concurrently request different content. Finally, as more detailed datasets become available, the Push-and-Track approach will be extended to include additional mobile participants such as pedestrians and opportunistic infrastructure (e.g., residential WiFi access-points or roadside units).

\section*{Acknowledgments}
We would like to especially thank the iTETRIS partners that have made available and built the vehicular dataset. For this, we especially thank Fabio Cartolano, Carlo Michelacci, and Antonio Pio Morra from the Municipality of Bologna, as well as Daniel Krajzewicz from the German Aerospace Center. We also thank Javier Gozalvez, Ramon Bauza, Cl\'emence Magnien, and Matthieu Latapy for their comments.
This work has been partly funded by the European project iTETRIS (No. FP7 224644) and the French ANR CROWD project under contract ANR-08-VERS-006.

\bibliographystyle{elsarticle-num}

\end{document}